\documentclass[12pt]{article}
\usepackage{amsmath}
\usepackage{floatflt}
\usepackage{amsfonts}
\usepackage{amssymb}
\usepackage{graphicx}
\usepackage{hyper}

\begin{document}

\title{The affine ambitwistor space as the moduli space of SUYM in $%
AdS_{5}\otimes S^{5}$}
\author{Bo-Yu Hou$^{a}$\thanks{%
Email:byhou@nwu.edu.cn}, Bo-Yuan Hou$^{b}$, Xiao-Hui Wang$^{a}$\thanks{%
Email:xhwang@nwu.edu.cn,} \\
Chuan-Hua Xiong$^{a}$, Rui-Hong Yue$^{a}$\thanks{%
Email:rhyue@nwu.edu.cn}\\
~\\
$^{a}$\textsl{Institute of Modern Physics, Northwest University,}\\
\textit{Xi'an, 710069, China}\\
$^b $\textsl{Graduate School, Chinese Academy of Science,}\\
\textit{Beijing 100039, China}}
\date{}
\maketitle

\begin{abstract}
By extending the dressing symmetric action of IIB string in $AdS_5\times S^5$
to the $D_3$ brane, we find a gauged WZW action of Higgs Yang-Mills field
including the 2-cocycle of axially anomaly. The left and right twistor
structure of left and right $\alpha $-planes glue into an ambitwistor. The
symmetry group of Nahm equations is central extended to an affine group,
thus we explain why the spectral curve is given by affine Toda.
\end{abstract}

\section{Introduction}

Bena, Polchinsky and Roiban \cite{bena} and Polyakov \cite{polyakov} show
the integrability of IIB string in $AdS_{5}\times S^{5}$. Using Metsaev
Tseytlin's action\cite{mt1}, we find in the $\kappa $ symmetric Killing
gauge, the $S$ duality between geometric Maurer Cartan equations and dynamic
equations of motion \cite{xiong1}. Then the reparametrization symmetry
permits a dual twisted transformation, so dictates the dressing symmetry.
The left and right moving world-sheet of string are chiral embedded in $%
AdS_{5}\times S^{5}$. By imposing the 2-cocycle of axial anomaly on Roiban
Siegel's action \cite{rs}, we get the gauged WZW action \cite{xiong}, which
dictates the conformal affine Toda model\cite{xiong}. We find the solitonic
string solutions, its moduli space realize the moduli of IIB string.

In this paper, we extended this to the super Yang-Mills theory in $D_{3}$
brane world sheet. At first, we shortly review the action for ADHM's anti
self-dual Yang-Mills and the action for Nahm-Drinfeld-Hitchin-Atiyah's
construction of BPS monopole. The twistor structure is manifest, since in
all the $\alpha $ plane, the anti self-dual Yang-Mills field becomes a pure
gauge. Next, from the Hamiltonian reduction induced by chiral vacuum
expectation value of Higgs field, we obtain a gauged WZW with ambitwistor
structure, the $TCP_{3}\times \overline{TCP}_{3}$ bundle. It happens that
the $D_{3}$ brane and IIB string shares a common covariant constant
quaternion 4-bein field embedded in the hyper-K\"{a}hler structure inherited
by $AdS_{5}$ background. This is exhibited by a common Robinson congruence,
its dilation and twist spin coefficients are given by the modulus and phase
of the level $\mu $ of hamiltonian reduction. So the zweibein of string is
uniquely completed to the 4-bein of brane. At last, we very shortly sketch
the original NDHA construction with a finite Toda type spectral curve and
suggest the way of affinization to a conformal affine Toda. And how the
moduli spaces of stretched string soliton corresponds to the moduli of
twisted monopole.

\section{Action of self-dual YM and twistor}

Firstly we consider the Lagrangian formulation of self-dual Yang-Mills
theory in four dimensional Euclidean space. We start from the $sp(4)$ gauge
field%
\begin{equation}
A_{\mu }=A_{\mu }^{a}t^{a}/2i,F_{\mu \nu }=F_{\mu \nu }^{a}t^{a}/2i=\partial
_{\mu }A_{\nu }-\partial _{\nu }A_{\mu }+\left[ A_{\mu },A_{\nu }\right] ,
\end{equation}%
where $t^{a}$ are Hermitian generators of sp(4). As in Yang's gauge \cite%
{yanggauge}, we adapt the complexified coordinates of Minkowski space%
\begin{align}
\sqrt{2}y& =x_{1}+ix_{2}\equiv x^{1\dot{1}};\sqrt{2}\bar{y}%
=x_{1}-ix_{2}\equiv x^{2\dot{2}};  \notag \\
\sqrt{2}z& =-x_{3}+ix_{4}\equiv x^{1\dot{2}};\sqrt{2}\bar{z}%
=x_{3}+ix_{4}\equiv x^{2\dot{1}}.
\end{align}%
and the metric becomes
\begin{equation}
ds^{2}=2dyd\bar{y}-2dzd\bar{z},
\end{equation}%
The K\"{a}hler form is self-dual (i.e. spans an $\alpha $-plane) and is
given by%
\begin{equation}
-2i\Omega _{1}=dy\wedge d\bar{y}-dz\wedge d\bar{z},
\end{equation}%
There are two other self-dual planes (i.e. $\alpha $-plane)
\begin{align}
\alpha & =dy\wedge dz=\Omega _{2}+i\Omega _{3},  \notag \\
\bar{\alpha}& =d\bar{y}\wedge d\bar{z}=\Omega _{2}-i\Omega _{3},
\end{align}%
and three anti-self-dual planes($\beta $-planes)
\begin{align}
\beta & =dy\wedge d\bar{z}=\bar{\Omega}_{2}-i\,\bar{\Omega}_{3}, \\
\bar{\beta}& =d\bar{y}\wedge dz=\bar{\Omega}_{2}+i\,\bar{\Omega}_{3}, \\
-2i\,\bar{\Omega}_{1}& =dy\wedge d\bar{y}-dz\wedge d\bar{z}.
\end{align}%
Then the equation of anti self-dual Yang-Mills
\begin{equation}
F_{\mu \nu }=\frac{1}{2}\varepsilon _{\mu \nu \alpha \beta }F_{\alpha \beta }
\end{equation}%
becomes
\begin{subequations}
\begin{align}
F_{yz}& =0,  \label{a} \\
F_{\bar{y}\bar{z}}& =0,  \label{c}
\end{align}%
\end{subequations}
\begin{equation}
F_{y\bar{y}}+F_{z\bar{z}}=0.  \label{b}
\end{equation}%
Since $\left( \ref{a}\right) $ and $\left( \ref{c}\right) $ implies
vanishing in $\alpha $ and $\bar{\alpha}$-plane, one may introduce two pure
gauge restricted to $\alpha $ and $\bar{\alpha}$-plane,
\begin{align}
A_{y}& =h^{-1}\partial _{y}h,A_{z}=h^{-1}\partial _{z}h  \notag \\
A_{\bar{y}}& =\bar{h}^{-1}\partial _{\bar{y}}\bar{h},A_{\bar{z}}=\bar{h}%
^{-1}\partial _{\bar{z}}\bar{h}
\end{align}%
where $h$ and $\bar{h}$ are $SL(4,c)$ matrix depending on $y,\bar{y},z$ and $%
\bar{z}$. Then $\left( \ref{a}\right) $ and $\left( \ref{c}\right) $ becomes
\begin{equation}
\left[ D_{y},D_{z}\right] =0,\left[ D_{\bar{y}},D_{\bar{z}}\right] =0.
\end{equation}%
For real $x,$ $A_{\mu }^{a}\doteq $ real gauge field, then%
\begin{equation}
\bar{h}\doteq \left( h^{\dagger }\right) ^{-1}.
\end{equation}%
here $\doteq $ imply restricted to real $x$. A gauge transformation is
characterized by the replacement%
\begin{equation}
h\rightarrow hg,\bar{h}\rightarrow \bar{h}g  \label{gaugetransformation}
\end{equation}%
with $g=g\left( y,\bar{y},z,\bar{z}\right) \in SL\left( 4,c\right)
,h^{\dagger }h\doteq \mathbb{I}=$ unit $4\times 4$ matrix, under which the
gauge potential $A_{\mu }$ and the field strength $F_{\mu \nu }$ transform as%
\begin{equation}
A_{\mu }\rightarrow g^{-1}\left( A_{\mu }+\partial _{\mu }\right) g,F_{\mu
\nu }\rightarrow g^{-1}F_{\mu \nu }g,
\end{equation}%
respectively. We introduce a Hermitian matrix%
\begin{equation}
U=h\bar{h}^{-1}\doteq hh^{\dagger },
\end{equation}%
which has the important property of being invariant under the gauge
transformation $\left( \ref{gaugetransformation}\right) $. Then $\left( \ref%
{b}\right) $ becomes
\begin{equation}
(U^{-1}U_{y})_{\bar{y}}+(U^{-1}U_{z})_{\bar{z}}=0,  \label{song}
\end{equation}%
where $U$ is $sp(4)$ matrix. It can also be written in a covariant form%
\begin{equation}
\bar{\partial}(U^{-1}\partial U)\wedge \Omega _{1}=0,  \label{d}
\end{equation}%
where
\begin{align}
\partial & =dy\partial _{y}+dz\partial _{z},  \notag \\
\bar{\partial}& =d\bar{y}\partial _{\bar{y}}+d\bar{z}\partial _{\bar{z}}.
\end{align}%
Hou and Song \cite{song}\ discover that Eq.$\left( \ref{b}\right) $
(equivalently to $(\ref{song}),(\ref{d})$) can be obtained from a variation
of the action,%
\begin{align}
S& =-\frac{4\pi }{2}\int_{E_{4}}Tr(dU^{-1}\wedge ^{\ast }dU)  \notag \\
& +\frac{4\pi }{3}\int_{M_{5}}Tr[(U^{-1}dU)^{3}]\wedge \Omega
_{1},~~\partial M_{5}=E_{4}  \label{songs}
\end{align}%
The Euler-Lagrange eq. of (\ref{songs}) is $\left( \ref{song}\right) .$ The
variation of this action is suggested by Donaldson in 1985 \cite{donaldson},
\begin{equation}
\delta S=-\int Tr(U^{-1}\delta U\partial (U^{-1}\bar{\partial}U))\wedge
\Omega _{1}.
\end{equation}%
Now we may combine $\left( \ref{a}\right) ,\left( \ref{c}\right) $ and $%
\left( \ref{b}\right) $ together%
\begin{align}
D_{1}& =D_{y}-\zeta D_{\bar{z}},  \label{operator} \\
D_{2}& =D_{z}-\zeta D_{\bar{y}},  \label{o1}
\end{align}%
where $\zeta $ is the CP(1) fibre of CP(3) bundle on base $S^{4}$ i.e.
compactified $E_{4}$, and they satisfy
\begin{equation}
\left[ D_{1},D_{2}\right] =0.  \label{flat}
\end{equation}%
When $\zeta =0,\infty $ and $1,$ (\ref{flat}) becomes (\ref{a}), (\ref{c})
and (\ref{b}) respectively. So in self-dual plane the anti-self-dual
connection is trivial, that is to say, they are sections of holomorphic
bundle on the self-dual plane
\begin{eqnarray*}
\Omega &=&(\Omega _{2}+i\Omega _{3})-2\Omega _{1}\zeta +(\Omega _{2}-i\Omega
_{3})\zeta ^{2} \\
&=&\alpha -2\Omega _{1}\zeta +\tilde{\alpha}\zeta ^{2}
\end{eqnarray*}%
here $\zeta $ describes the $C\mathbb{P}_{1}$ set of complex structures.

\section{Dressing symmetric action of Higgs YM fields}

The D3-brane moving in $AdS_{5}\otimes S^{5}$ background should be described
by DBI action as by Metsaev and Tseytlin \cite{mtd3}. But as point out by
Kallosh and Rajaraman \cite{kallosh9805041} on the maximally supersymmetric $%
AdS_{5}\otimes S^{5}$ background, the basic scalar superfield are covariant
constant and the fermionic superfield $\Lambda _{\alpha }$ (related with the
S-duality $U(1)\subset \frac{SU(1,1)}{U(1)}$) vanishes. So in $\kappa $
symmetric static Killing gauge, the background is pure geometrical. The only
field remains essentially are the $R_{ab}^{\quad cd}\sim \eta _{a}^{c}\eta
_{b}^{d}-\eta _{a}^{d}\eta _{b}^{c}$ and 5-form field $\sim $ $\epsilon
_{abcde}$, all others can be expressed by the geometrical quantity such as
torsion. Thus Kallosh and Rajaraman \cite{kallosh9805041} argued that
D3-brane may be simply written as the nonlinear sigma models. To display the
hidden symmetry of the moduli, it is sufficient to restrict to the bosonic
case. So to describe the chiral embedding of $D_{3}$ brane in $%
AdS_{5}\otimes S^{5}$, we will construct the nonlinear sigma models by
gauging\ the WZW form of last section. But now instead of Euclidean world
sheet $E_{4},$ we will change to Minkowski world sheet $M_{4}\left(
t,x_{1},x_{2},x_{3}\right) $ with topology $S^{1}\times \mathbb{R}^{3}$. Now
we define%
\begin{align}
\sqrt{2}y& =t+x_{1};\sqrt{2}\bar{y}=t-x_{1};  \notag \\
\sqrt{2}z& =-x_{2}+ix_{3};\sqrt{2}\bar{z}=x_{2}+ix_{3}.
\end{align}

The $SU(2,2)$ symmetry is extended to $GL(4)^{(1)}$ as \cite{xiong}. Then
adding the chiral anomaly 2 cocycles term, the action is given as%
\begin{align}
S& =-\frac{4\pi }{2}\int_{M_{4}}Tr(dU^{-1}\wedge ^{\ast }dU)  \notag \\
& +\frac{4\pi }{3}\int_{M_{5}}Tr[(U^{-1}dU)^{3}]\wedge \Omega _{1}  \notag \\
& +\frac{4\pi }{2}\int_{M_{4}}d^{2}yd^{2}zTr\{A_{y}\left( U^{-1}\partial _{%
\bar{y}}U-\mbox{\boldmath $\mu$}\right) +\left( \partial _{y}UU^{-1}-%
\mbox{\boldmath $\nu$}\right) \tilde{A}_{\bar{y}}+\tilde{A}_{\bar{y}%
}UA_{y}U^{-1}\}  \notag \\
& -\frac{4\pi }{2}\int_{M_{4}}d^{2}yd^{2}zTr\{A_{z}\left( U^{-1}\partial _{%
\bar{z}}U-\mbox{\boldmath $\mu$}\right) +\left( \partial _{z}UU^{-1}-%
\mbox{\boldmath
$\nu$}\right) \tilde{A}_{\bar{z}}+\tilde{A}_{\bar{z}}UA_{z}U^{-1}\}.
\label{gaugewznw}
\end{align}%
here
\begin{equation}
\mbox{\boldmath $\mu$}=\mu E_{+}\equiv \mu \left(
\begin{array}{cccc}
0 & 1 &  &  \\
& 0 & 1 &  \\
&  & 0 & 1 \\
1 &  &  & 0%
\end{array}%
\right) ,\mbox{\boldmath $\nu$}=\nu E_{-}\equiv \nu \left(
\begin{array}{cccc}
0 &  &  & 1 \\
1 & 0 &  &  \\
& 1 & 0 &  \\
&  & 1 & 0%
\end{array}%
\right)\label{f}
\end{equation}%
in principal representation. Later we will always choose $\mu =\nu ^{-1},$
so the $4\pi $ before 3rd and 4th integral is the mass \cite%
{atiyahandhitchin} of monopole.\ Here if we consider SP(4) as $SL_{2}(H)$,\
then the left $A$ correspond to the algebra of left multiplication group of
quaternion, the $\tilde{A}$\ to the right conjugate one. So left $\alpha $%
-plane correspond to that of the twistor $Z^{a}$ and right $\bar{\alpha}$%
-plane for the anti-twistor googly $\bar{Z}^{a}.$ Thus the model are section
of ambitwistor bundle with base $\mathbb{C}M_{4}$ and the two left right $%
\mathbb{C}P_{1}$ fibre are mutually conjugate and trivialized by $\zeta $
and $\bar{\zeta}$ respectively.\ Varying the action with respect to $%
A_{1}=A_{y}-\zeta A_{\bar{z}},A_{2}=A_{z}-\zeta A_{\bar{y}},$ etc. and $%
\delta U=\epsilon _{L}U,\delta U=U\epsilon _{R}$ respectively, we obtain the
equation of motion of $A,\tilde{A}$ and $U$,
\begin{align}
& (\partial (U^{-1}\tilde{\partial}U+U^{-1}\tilde{A}U)-[A,U^{-1}\tilde{%
\partial}U+U^{-1}\tilde{A}U]+\tilde{\partial}A)\wedge \Omega _{1}=0 \\
& (\tilde{\partial}(\partial UU^{-1}+UAU^{-1})-[\tilde{A},\partial
UU^{-1}+UAU^{-1}]+\partial A)\wedge \Omega _{1}=0 \\
& Tr[E_{\alpha _{+}}(U^{-1}\partial _{\bar{y}}U+U^{-1}\tilde{A}_{\bar{y}}U-%
\boldmath{\nu })]=0  \label{31} \\
& Tr[E_{\alpha _{+}}(U^{-1}\partial _{\bar{z}}U+U^{-1}\tilde{A}_{\bar{z}}U-%
\boldmath{\nu })]=0 \\
& Tr[E_{\alpha _{-}}(\partial _{y}UU^{-1}+UA_{y}U^{-1}-\boldmath{\mu })]=0 \\
& Tr[E_{\alpha _{-}}(\partial _{z}UU^{-1}+UA_{z}U^{-1}-\boldmath{\mu })]=0
\label{34}
\end{align}%
Here $U$\ is the transfer matrix in two different order of Gauss
decomposition (cf. \cite{xiong}). After acting on the highest (lowest)
weight vector of level one affine algebra, only the upper and lower Borel
part remain and we will get sections of holomorphic and anti-holomorphic
bundle respectively.

\section{Nahm Donaldson Hitchin Atiyah's monopole moduli}

In next section, we will argue that the gauged WZW action (\ref{gaugewznw})
are the axially twisted affinize generalization of Nahm, Donaldson, Hitchin,
Atiyah's theory of monopole. The holomorphic structure of the left and the
right anti-holomorphic structure are glued in an axially twisted way.

Nahm's formalism for the BPS monopoles is obtained by Fourier transform the
ADHM construction of multi-instantons to a Hilbert space.

\subsection{ADHM construction of instanton\protect\cite{adhm,adhmatiyah}}

In su(n) self-dual Yang-Mills theory, the potential is
\begin{equation}
A_{\mu }=\upsilon ^{\dagger }\partial _{\mu }\upsilon ,  \label{1}
\end{equation}%
\begin{equation}
\upsilon ^{\dagger }\upsilon =1,  \label{2}
\end{equation}%
here $\upsilon $ is $\left( 2k+n\right) \times \left( 2k\right) $ matrices
with instanton number $k$. Properly choose a $\left( 2k+n\right) \times
\left( 2k\right) $ matrices%
\begin{equation}
\bigtriangleup (x)=a+b\mathbf{x},
\end{equation}%
\begin{equation}
\mathbf{x}=x_{\mu }q^{\mu },
\end{equation}%
where $\mathrm{q}$ is quaternion. Then the solution $\upsilon $\ of
\begin{equation}
\bigtriangleup ^{\dagger }\upsilon =0,  \label{39}
\end{equation}%
and (\ref{2}) give the potential (\ref{1}).

\subsection{Nahm equation \protect\cite{nahm}}

For the su(2) BPS monopole with magnetic charge $k$, Nahm introduce a $%
2k\times 2$ matrix function $\upsilon (s)$ of $s$ satisfying
\begin{align}
\left( i\frac{d}{ds}+x^{\dagger }+T^{\dagger }\right) \upsilon & =0,  \notag
\\
\int\nolimits_{0}^{2}\upsilon ^{\dagger }\upsilon ds& =1.
\end{align}%
where $x^{\dagger }=x^{i}q^{i}~~(i=1,2,3)$ and $s$ is the Fourier
transformation of $x_{0}$. Nahm has show that the matrix $T=T_{\mu }q^{\mu }$
satisfy the self-dual Nahm equation (equivalent to Donaldson's \cite%
{donaldson96} complex representation)%
\begin{align}
\frac{dT_{i}}{ds}& =\frac{1}{2}\sum \epsilon _{ijk}\left[ T_{j},T_{k}\right]
+\left[ T_{0},T_{i}\right] ,  \notag \\
T_{\mu }^{\dagger }& =\eta _{\mu \nu }T_{\nu }.  \label{s0}
\end{align}%
The $T_{0}$ can be gauge transformed into zero, by $g(s)$ with $g(s)\overset{%
s\rightarrow 0}{\longrightarrow }1,$ at last we have the Nahm equation,%
\begin{equation}
\frac{dT_{i}}{ds}+\frac{1}{2}\sum \epsilon _{ijk}\left[ T_{j},T_{k}\right]
=0,  \label{s1}
\end{equation}%
\begin{equation}
T_{i}^{\ast }(s)=-T_{i}(s),
\end{equation}%
\begin{equation}
T_{i}(2-s)=T_{i}(s)^{T},
\end{equation}%
\begin{equation}
\text{The }T^{i}\ \text{are analytic over }(0,2),\text{with simple poles at }%
0,2.
\end{equation}%
\begin{equation}
\text{The residues of }T_{i}\text{ at }s=0\text{ form an irreducible
representation of }su(2).  \label{s5}
\end{equation}

\subsection{The Donaldson's rational map and Hitchin's spectral curve}

Donaldson has established a one to one correspondence between the
equivalence class of Nahm's complex (\ref{s1})-(\ref{s5}) and a rational
function $S(z)$. Donaldson's $S(z)$ is a rational function of degree $k,$
regular at $\infty ,$ $S(\infty )=0$.
\begin{equation}
S(z)=<s|^{T}(zI-B)^{-1}|s>~~\in \mathbb{C}\cup {\infty }  \label{47}
\end{equation}%
where $(i)$ $B$ is a symmetric $k\times k$ matrix and $|s>\in \mathbb{C}^{k}$
is a column vector.\newline
$(ii)$ $|s>$ generates $\mathbb{C}^{k}$ as a $\mathbb{C}[B]$ module.

Hitchin established the correspondence between the holomorphic vector bundle
and the solution of BPS eq. and explain the $S(z)$ as scattering function.
He consider the differential operator
\begin{equation}
\nabla _{U}-i\phi ,\nabla _{i}=\partial _{i}+iA_{i},A,\phi \text{ solution
of BPS eq.}  \label{48}
\end{equation}%
acting on section $s(t)\in TP_{1}(C)$ over a fixed oriented line $U$ with
parameter $t$. If there is a solution $s_{0}$ decays at both $t\rightarrow
+\infty $ and $t\rightarrow -\infty .$ A line $U$ with this property then is
called in \cite{atiyahandhitchin} a spectral line. Thus the poles of $S(z)$
represent the $k$ spectral lines parallel to the $x_{1}$-axis passing
through the $x_{1}=0$\ plane at $z=x_{2}+ix_{3}.$ The space of all oriented
straight lines in $\mathbb{R}^{3}$ has identified with the tangent bundle of
the complex projective line $T\mathbb{P}_{1}$\cite{hitchincmp83}. The
subspace consisting of all spectral lines is called the spectral curve $S$
\cite{hitchincmp89}.

\section{The Ambitwistor description of moduli space}

\subsection{Mini-twistor $\mathbb{TP}_{1}$ and Robinson congruence of $TCP_3
\times \overline{TCP_3}$}

Using the Jacobi vector field, Hitchin \cite{hitchincmp83} define the
complex structure for the space $TP_{1}$ of the oriented straight lines in $%
\mathbb{R}^{3}.$ The real structure $\tau $ is given by reversing the
orientation. The conformal structure is induced on $\mathbb{R}^{3}$ with the
metric of 3 dimensional Euclidean space $E_{3}.$ Hitchin consider the
geometry of the complexification $\mathbb{C}^{3}$ of $\mathbb{R}^{3}$ define
a mini-twistor representation. Mini-twistor is the quotient space of $%
CP_{3}/CP_{1}$ by the action of time translation which is the real part of
the action of $C$. A point $(a,b,c)$ in $\mathbb{C}^{3}$ is represented by a
holomorphic action $(a\zeta ^{2}+b\zeta +c)\frac{d}{d\zeta }$ of $T\mathbb{P}%
_{1}$. The sections through a fixed point $(\eta _{0},\zeta _{0})\in T%
\mathbb{P}_{1}$. $\{(a,b,c)\in C^{3}|\eta _{0}=a\zeta _{0}^{2}+b\zeta
_{0}+c\}$ is a null plane in $C^{3}$. His null planes give the usual null
geodesic congruence of real $M_{4}$ \cite{penrose}.

Now in our action (\ref{gaugewznw}) restricted to the null cone component $%
dy=\frac{dt+dx_{1}}{\sqrt{2}},\bar{dy}=\frac{dt-dx_{1}}{\sqrt{2}}$, it turns
to be the same as the affine Toda case in \cite{xiong}. Here in D3 brane
case, the left moving and right moving world-sheet has been extended to $%
\mathbb{CM}_{4},$ So the zweibein $dy,d\bar{y}$ (the corresponding tangent
vectors are null momentum) corresponds to spinor $\lambda ^{\alpha },\bar{%
\lambda}^{\dot{\alpha}},$ and can be completed to the full 4-bein
corresponding to the twistor%
\begin{equation*}
X=\left( \lambda ^{\alpha },\mu _{\dot{\alpha}}\right) \text{ and }\bar{X}%
=\left( \mu ^{\alpha },\lambda _{\dot{\alpha}}\right) ,
\end{equation*}%
which is the Robinson congruence, that is it intersect with a non-null
twistor Z%
\begin{equation*}
\bar{Z}_{a}X^{a}=0\text{ with }\bar{Z}_{a}Z^{a}=m^{2}\neq 0.
\end{equation*}

Now we will explain why dressing symmetric string embedding and $D_{3}$
brane embedding are described by the same Robinson congruence. Both string
and $D_{3}$ brane share the same quaternion basis inherited by the target
space $AdS_{5}\sim \frac{GL(4)}{Sp(4)\otimes U(1)}$. The Serret Frenet
equation of complex curve spanned by the string world sheet implies a
covariant invariant Jacobi field with covariant constant quaternion
basis--the moving frame. Since the embedding of $E_{\pm }$ in $AdS_{5}$ is
principal, so the moving of zweibein of tangent vector will determine a
complete 4-bein by the normal, binormal etc of this complex curve, this will
be the same covariant constant Jacobi field for the moving frame" of
quaternion curve", which describes the embedding of brane. That is the same
Robinson congruence with the same parameter $\mu =\nu ^{-1}$ (m has been
fixed). It is crucial that the Robinson congruence is shearfree \cite%
{penrose}. So $|\mu |=$ the dilation spin coefficient and $\frac{\mu }{|\mu |%
}=$ the twist uniquely determine and is determined by the common dilation
and rotation of 3 space component for the Jocabi field. From the quaternion $%
\eta $\ and Jocabi field, one may construct a hyper k\"{a}hler structure
\cite{atiyahandhitchin} which is covariant constant \cite{atiyahandhitchin}.
These constitute the 3 family of $\alpha $\ planes. So from Robinson
congruence of one set $\alpha $\ planes, one may get the whole hyper k\"{a}%
hler structure. The Ward transition function $U$\ and factorized $W$ \cite%
{wittentwistor}\ may be constructed also. Now the W is realized by the
Riemann Hilbert transformation of affine Toda.

\subsection{NDHA moduli of monopole-----Toda; moduli of dressing affinized
case-----affine Toda}

Now we turn to argue that as the \textbf{NDHA's moduli space} is described
by spectral curve of \textbf{Toda}, the \textbf{moduli space of gauged WZW} (%
\ref{gaugewznw}) is described by spectral curve of \textbf{affine Toda}. The
group $GL(4)^{(1)}$ in action (\ref{39}) is the central extension of the $%
SL(4)$ in action (\ref{songs}). The Lie algebra structure in Nahm's eq. will
be replaced by an affine one. The section $|s\rangle \in L^{2}\times
\mathcal{O}\left( k\right) $ in the kernel of left $\bigtriangleup $
operator and section $\langle s|$ in the kernel of right $\bigtriangleup
^{\ast }$operator will be replaced by the heighest weight space vector $|%
\hat{s}\rangle $ and $\langle \hat{s}|$ of affine algebra. The Toda type
\cite{gervais}
\begin{equation}
U^{-1}\partial _{+}U=\left(
\begin{array}{cccc}
0 & \cdots & \cdots & 0 \\
1 & \ddots &  &  \\
& \ddots & \ddots &  \\
&  & 1 & 0%
\end{array}%
\right) ~~~~~~~U^{-1}\partial _{-}U=\left(
\begin{array}{cccc}
0 & 1 &  & 0 \\
& \ddots & \ddots &  \\
&  & \ddots & 1 \\
&  &  & 0%
\end{array}%
\right)
\end{equation}%
will be replaced by the affine one (\ref{f}). Let
us illustrate these modification in the following equivalence class of \cite%
{hitchincmp89}.\newline

\textbf{A}. A solution to the Bogomolny equations $D\Phi =\ast F$ on $%
\mathbb{R}^{3}$ with boundary conditions as $r\rightarrow \infty ,$

\begin{enumerate}
\item $||\Phi ||=1-\frac{k}{r}+O\left( r^{-2}\right) ,$

\item $\frac{\partial ||\Phi ||}{\partial \Omega }=O\left( r^{-2}\right) $,

\item $||D\Phi ||=O\left( r^{-2}\right) $.
\end{enumerate}

\textbf{B}. A compact algebraic curve $S\in T\mathbb{P}_{1}$ in the linear
system $|\mathcal{O}(2k)|.$

\textbf{C}. The Nahm complex eq.(\ref{s1})-(\ref{s5})

\subsection{Spectral curve $(\mathbf{B}\rightarrow \mathbf{C})$}

As in Hitchin \cite{hitchincmp83} theorem (6.3) and P156 in \cite%
{hitchincmp89}, let $\tilde{E}$ be the holomorphic bundle on $T\mathbb{P}%
_{1} $ represented by the exact sequence%
\begin{equation*}
0\rightarrow L\left( -k\right) \rightarrow \tilde{E}\rightarrow L^{\ast
}\left( k\right) \rightarrow 0,
\end{equation*}%
There are two distinguished bundle $L^{+}$ and $L^{-}$
\begin{equation*}
L^{\pm }=\{u\in \tilde{E}|u\left( t\right) \rightarrow 0,t=\pm \infty \}
\end{equation*}%
where $L^{+}\simeq L(-k),L^{-}$ are the conjugate of $L^{+}$ by real
structure. the spectral curve S is defined by%
\begin{equation*}
S=\{u\in T\mathbb{P}_{1}|L_{u}^{+}=L_{u}^{-}\}.
\end{equation*}%
Then as in proposition (7.3) in \cite{hitchincmp83} and P156 in \cite%
{hitchincmp89}, S is the divisor of a section of $\psi \in H\left( T\mathbb{P%
}_{1},\mathcal{O}\left( 2k\right) \right) $ which is defined by the equation%
\begin{equation}
S=\{(\eta ,\zeta )\in T\mathbb{P}_{1}|\psi =\eta ^{k}+a_{1}\left( \zeta
\right) \eta ^{k-1}+\cdots +a_{k}\left( \zeta \right) =0\},  \label{s50}
\end{equation}%
where $a_{i}\left( \zeta \right) $ is a polynomial of degree $2i$ in $\zeta $%
.

And this spectral curve has been identified as p170 in \cite{hitchincmp89}%
\begin{eqnarray}
S &=&\{\left( \eta ,\zeta \right) \in T\mathbb{P}_{1}|\det \left( \eta
+A\left( \zeta \right) \right) =0\}, \\
A &=&A_{0}+\zeta A_{1}+\zeta ^{2}A_{2}.
\end{eqnarray}%
As shown by \cite{hitchincmp89}, the equation (\ref{48}) is the spectral
curve of \textquotedblleft open chain\textquotedblright\ Toda, $detL_{Toda}=0
$.\footnote{%
For example Krichever and Vanisky \cite{krichever} show that the spectral
curve of open Toda chain is that as in (\ref{s50}) with $\psi \rightarrow
\xi $\ . Marshakov \cite{marshakov} prove it is the perturbative limit of
the affine Toda type spectral curve of quantum moduli space for the 4D pure
SUYM.} This can be explained by Gauss decomposition of the $U$ in WZW action
(\ref{songs}). As in \cite{hurtubise97}, the upper and lower triangle part
becomes holomorphic and anti holomorphic respectively, thus corresponds to $%
L^{+}$ and $L^{-}$.\newline

After find the independent of the eigenvector $\eta $, Hintchin shows the
Toda type Lax pair,%
\begin{equation}
\left[ A_{+},A\right] +\frac{dA}{ds}=0,  \label{53}
\end{equation}%
$A_{+}\equiv \frac{1}{2}A_{1}+\zeta A_{2}$. Let $%
A_{0}=T_{1}+iT_{2},A_{1}=2iT_{3},A_{2}=T_{1}-iT_{2},$ equation (\ref{53})
turn to be the Nahm equation.

In the process of determine the surface S by the spectral eq. the essential
point is to use the series\ of exact sequence of inclusion $i$, restriction $%
\rho $ and the coboundary $\delta $. In the ambitwistor case, this $\delta $%
\ of Cech double cohomology will be given by chiral anomaly as shown by \cite%
{houwang,hhbook}, realized in our affine action (\ref{gaugewznw}) where $U$
turns to be an element of the affine group.

The $\hat{S}$ becomes
\begin{equation*}
\hat{S}=\{u\in \mathbb{CP}_{3}\times \overline{\mathbb{CP}_{3}^{\ast }}%
|L_{u}^{+}=(L_{u}^{-})^{\ast }\},
\end{equation*}%
where $L^{-}=e^{k_{-}}M_{-},L^{+}=e^{k_{+}}M_{+}$ is the upper (lower)
triangle part in two different order of Gauss decomposition $%
U=e^{k_{-}}N_{+}M_{-}=e^{k_{+}}N_{-}M_{+}$, with $k_{\pm }$ diagonal, $%
N_{+},M_{+}$ upper triangle and $N_{-},M_{-}$ lower triangle. So $\hat{S}$
is hyper-elliptic, symmetrical under $\zeta \rightarrow \frac{1}{\zeta }$.
\begin{equation*}
\hat{S}=\{(\eta ,\zeta ):\zeta +\frac{1}{\zeta }=\eta ^{k}+a_{1}\eta
^{k-1}+\cdots +a_{k}=0\},
\end{equation*}%
which is the spectral curve of affine Toda. For affine Toda case, $\hat{L}$
is determined by the vanishing of double pole of Akhizer-Baker operator, $%
\hat{A}_{+}$ is by vanishing of the simple pole , which is independent of $%
\eta $ by the dispersion phase of AB function. So in the affine Toda case,
Nahm equation is satisfied also.

\subsection{The boundary behaviour of Higgs $\Phi $ of BPS equation and Nahm
equation ($\mathbf{C}\longleftrightarrow \mathbf{A}$)}

The $\Phi (\infty )$ up to $\frac{1}{r^{2}}\in $ Sobolev space $H^{\prime }$%
, is given by coboundary, which is determined by the residue of Nahm's $%
T_{i}(s)$ at $s=0$. Hitchin show \textbf{A} by consider the pole of $\langle
$ kernel of $\Delta ^{\ast }$, kernel of $\Delta \rangle $ , Hurtbise \cite%
{hurtubise97} find the coboundary by using Lagrange interpolation, to get
the transition matrix, so that the $\frac{1}{r^{2}}$ order is given by the
level $\mu $, which describes the central position of the monopole, i.e. the
scale of the Jacobi field, given by solution of the linearized BPS equation
\cite{hitchincmp83,atiyahandhitchin}.

In affine Toda case, we know from the action (\ref{gaugewznw}), the $\mu
\sim <\Phi >\longleftrightarrow \Phi (\infty )$. Now the uaual $SL(2)$
algebra has been complexified, So $<\Phi >$ will be complex.

The kernel of affine $\hat{\Delta},\hat{\Delta}^{\ast }$ is given by affine
vertex operator $V(\xi )$ with pole and zero at $\zeta =\mu $. Here $\mu $
is the rapidity, i.e. the scale and phase of the Jacobi field along the null
direction. It is the common scale of all three space components, three
Jacobi fields. This is in consistent with the well known fact, that the
radial and transverse component of Higgs and Yang-Mills filed satisfying BPS
equation will give asymptotic value as \textbf{A} (e.g. \cite%
{atiyahandhitchin} (4.5)), the radial part gives the monopole value $k$\ and
scale $\mu $.

It worth mention that the quantum $k$, the topological charge of S.G.
contributed by boundary flow is extended to the charge of monopoles, the
magnetic flux which actually realized the coboundary, given by the 1st Chern
class.

\subsection{Donaldson's classification}

Hurtbise \cite{hurtubise100} has chosen an trivialization for the solution $%
s(t)$ of Nahm equation. Such that both $s(0)$ and $s(1)$ becomes $\cong
(1,1,\cdots ,1)$. Then as Donaldson (\ref{47}), the residue of $B(0)$
becomes $\left(
\begin{array}{cccc}
0 &  &  &  \\
1 & 0 &  &  \\
& \ddots & \ddots &  \\
&  & 1 & 0%
\end{array}%
\right) .$

In the affine case, corresponding to the constrain (\ref{47}), the residue
should be replaced by the cyclic element in affine algebra. \cite{kac} $%
\left(
\begin{array}{cccc}
&  &  & \mu \\
\mu &  &  &  \\
& \mu &  &  \\
&  & \mu &
\end{array}%
\right) $ acting as permutations on left highest weight space and $\left(
\begin{array}{cccc}
& \mu ^{-1} &  &  \\
&  & \mu ^{-1} &  \\
&  &  & \mu ^{-1} \\
\mu ^{-1} &  &  &
\end{array}%
\right) $ acting on right conjugate space. This is really the behaviour at
the multiple pole and zero at $\zeta \longrightarrow \pm \infty $ on
spectral curve of affine Toda \cite{marshalov9802007,kcmp}.

\subsection{Solution space of twisted monopole}

The utility of gauged WZW with affinization is that it implies dressing
transformation and Riemann Hilbert problem. Thus we may find soliton
(twisted monopole) solution and describe the moduli space by $\mu $ as
following.

The affine algebra with dressing symmetry will enable us to find the soliton
solution. One may find the stretched IIB string solution for the conformal
affine Toda \cite{babelon,olive}. This turns to be the left and right ray in
the Robinson congruence. Meantime the other two orthogonal component, the
Jacobi field is given also \cite{prani,sachs}, as in the Robinson
congruence. So it determines the four dimension monopole solution. Here, it
is important that to obtain an soliton solution we have start from the
highest weight vector with nontrivial dependence of $z_{+},z_{-}$\ (cf \cite%
{babelon}), then acted by the left $g^{+}$ and right $g^{-}$\ part of the
dressing group $g$, which can be expressed by the $\pm $ frequency part of
vertex operator of affine algebra. To obtain multi-soliton, one may simply
use the Wick theorem of normal order. The Riemann Hilbert problem with zero
and pole give the classification of moduli space, implies implicitly the
commutative rule of multi Backlund transformation.

It should stressed that the dressing symmetry, the Poisson Lie structure of
the bi-algebra are important, since the classical $r$ matrix can be
quantized to quantum $R$ matrix. The quantized Yang-Baxter representation
operator $L$\ and vertex operators (type I and type II) has been given. And
the quantum affine algebra, its currents and q-deformed Virasoro
(W-algebra.) is given, including the affine Toda and sine Gordan. This may
provide the formulation of covariant quantization of string and SUYM
explicitly in the Green-Schwarz formulation.

\subsection{SUYM and brane}

The level $\mu $\ in hyper k\"{a}hler quotient are realized as the vacuum
expectation value $\langle \phi \rangle ,$ usually in the following way

\begin{enumerate}
\item Appears in D term \cite{Douglas9603167} as FI term and superpotential
associate with anomaly. Then for brane set D3-D1, N5-D3, the equation of
motion becomes the Nahm eq. or the Hitchin system. For example \cite%
{marshakov,diaconesca,hannywitten9611230,witten9703166}.

\item The Seiberg Witten curve is related with integrable system. For pure
SU(N) super Yang-Mills theory, it is given by the spectrum curve of affine
Toda.
\end{enumerate}

It is plausible that this two fact will unified by the hidden symmetry in
gauged WZW embedding in twistor space $C\mathbb{P}_{3},$ further to $C%
\mathbb{P}^{3|4}.$ This implies we may simply affinize the Nahm eq. with an
axial definition of conjugation. Then the vacuum moduli may be classified by
the moduli space of its soliton monopole solution, the twisted monopole and
the Seiberg Witten curve of the moduli of SUYM will be identified with the
spectral curve $|\eta +L\left( \zeta \right) |.$

In summary, the expert will find that the added term couple to $A$ are the
familiar $2$ cocycle of axial anomaly i.e. D term and FI term \cite%
{Douglas9603167,witten9301042}, so the level \ $\mu $\ equals the vacuum
eigenvalues of Higgs field in the $N=1$ chiral multiplet of $N=2$ vector
multiplet,%
\begin{equation}
\mu =\langle \phi \rangle .
\end{equation}%
Really as shown by Diaconescu \cite{diaconesca} the $\mu $\ will be the VEV
of \ $\phi $\ or its asymptotic value. Our twist of the parameter $\mu $\
will change the module and phase of it which correspond to dilation of
monopole and dual rotation of electric and magnetic (topologic) charge, or
equivalently change the phase $t$ and $r$ of the $\theta $\ term and $D$
term \cite{witten9301042}, which is the charge of Hodge structure of the A,
B twist as Witten describes in paper \cite{mirro}. All are related to S
duality i.e. $U(1)$ in the $\frac{SL(2R)}{U(1)}$ of SUGRA and M theory which
is restricted to the $SL(2Z)$ by Dirac quantization. Now we have continuous $%
SL(2R)$ to describe different point (vacuum) in moduli space, where the
phase of $\langle \phi \rangle $, just implies a redefinition of the $U(1)$
charge (\textquotedblleft electric\textquotedblright ) minimally coupled to $%
\phi .$ It would be interest to study the continuous effect of this affine
gl(1C) such as: 1. the $U(1)$ current in CFT, related to the spectral flow
for chiral ring \cite{warner}; 2. the special geometry \cite{specialg}; 3.
the Hanny Witten effect \cite{hannywitten9611230}; 4. the blow up for A
brane to B brane \cite{vafablowup}. We expect that this hidden symmetry will
be helpful to clarify the singularity, the cut of moduli space, the phase
diagram of super Yang-Mills theory.

Recently there are a lot of papers on (super)twistor formulation of
superstring \cite{wittentwistor}. It would be interest to study its dressing
symmetric form with affine symmetry $gl(4|4)^{\left( 1\right) }$ and to
study the CFT/AdS correspondence, by using the quantum double covariant
quantization of the classical double ambitwistor.

\section*{Acknowledgments}

The authors thank Kang-Jie Shi for useful discussions.

\noindent We also thanks Yong-Shi Wu, who informs us that the paper \cite%
{0407044} further gives the super Yangian form for Dolan Nappi Witten's
version of integrability \cite{witten,dw04}.

\section*{Note}

This paper gives the conjecture for the affinization of NADHM
construction. Recently, we find a left (right) moving ,
holomorphic (anti), selfdual(anti) spherical symmetrical solution
for affine Higgs Yang-Mills field, and the Dirac eq. on this
background. Thus realize the NADHM construction explicitly.

\end{document}